# A SIMPLE 1-BYTE 1-CLOCK RC4 HARDWARE DESIGN AND ITS IMPLEMENTATION IN FPGA COPROCESSOR FOR SECURED ETHERNET COMMUNICATION

Rourab Paul[1], Sangeet Saha[2], J K M Sadique Uz Zaman[3], Suman Das[4],
Amlan Chakrabarti[5] and Ranjan Ghosh[6]

Email :{rourabpaul[1], sangeet.saha87[2], jkmsadique[3], aami.suman[4]}@gmail.com,
acakcs@caluniv.ac.in[5], rghosh47@yahoo.co.in[6]

A.K.Choudhury School of Information Technology[1,5], Dept. of Computer Science and Engineering[2] and
Institute of Radio Physics and Electronics[3,4,6],
University of Calcutta, 92 A. P. C. Road, Kolkata – 700 009, India

## Abstract

In the field of cryptography till date the 1-byte in 1-clock is the best known RC4 hardware design [1], while the 1-byte in 3clocks is the best known implementation [2,3]. The design algorithm in [1] considers two consecutive bytes together and processes them in 2 clocks. The design of 1-byte in 3-clocks is too much modular and clock hungry. In this paper considering the RC4 algorithm, as it is, a simpler RC4 hardware design providing higher throughput is proposed in which 1-byte is processed in 1-clock. In the design two sequential tasks are executed as two independent events during rising and falling edges of the same clock and the swapping is directly executed using a MUX-DEMUX combination. The power consumed in behavioral and structural designs of RC4 are estimated and a power optimization technique is proposed. The NIST statistical test suite is run on RC4 key streams in order to know its randomness property. The encryption and decryption designs are respectively embedded on two FPGA boards with RC4 in a custom coprocessor followed by Ethernet communication.

## 1. Introduction

RC4 is a widely used stream cipher whose algorithm is very simple. It has withstood the test of time in spite of its simplicity. The RC4 was proposed by Ron Rivest in 1987 for RSA Data Security and was kept as trade secret till 1994 when it was leaked out [4]. Today RC4 is a part of many network protocols, e.g. SSL, TLS, WEP, WPA and many others. There were many cryptanalysis to look into its key weaknesses [4, 5] followed by many new stream ciphers [6, 7]. RC4 is still the popular stream cipher since it is executed fast and provides high security.

There exist hardware implementations of some of the stream ciphers in the literature [8-11]. Since about 2003 when FPGA technology has been matured to provide cost effective solutions, many researchers started hardware implementation of RC4 as a natural fall out [2,3]. The FPGA technology turns out to be attractive since it provides soft core processor having design specific functional capability of a main processor (MicroBlaze [12]) along with reconfigurable logic blocks that can be synthesized to a desired custom coprocessor, embedded memories and IP cores. One can design RC4 algorithm totally as an executable code for the soft core processor (main processor) only or in custom coprocessor hardware operated by the main processor. Because of the system overhead, any single instruction if executed in the main processor takes at least 3 clocks, while the identical one when executed in a coprocessor takes 1-clock as the latter is customized to handle the specific task. Besides the clock advantage, the coprocessor based design makes the system throughput faster by another fold since it is executed in parallel with the main processor.

In this paper, RC4 algorithm is considered as it is and exploiting conventional VHDL features a design methodology is proposed processing of 1-byte in 1-clock. The said design is implemented in a custom coprocessor functioning in parallel with a main processor (Xilinx Spartan3E XC3S500e-FG320 FPGA architecture) followed by secured data communication between two FPGA boards through their respective Ethernet ports – each of the two boards performs RC4 encryption and decryption engines separately. The performance of our design in terms of number of clocks proved to be better than the previous works [1 – 3]. The clock gating technology is introduced to save dynamic power. In order to see the resilience of RC4, a battery of statistical tests as mentioned in the NIST document [14] is undertaken and it is found that the randomness property of its key streams is reasonably good.

The paper is organized as follows. The RC4 algorithm is briefly described in Sec. 2. In Sec. 3 the 1-byte-1-clock design and its hardware implementations are described. The communication experiment set up along with the results of relative comparisons is narrated in Sec. 4. The power optimization using clock gating technology is discussed in Sec.5. The randomness test of RC4 algorithm undertaken following the NIST statistical tests suite is discussed in Sec. 6 along with its results. The conclusion is discussed in Sec. 7.

## 2. RC4 Algorithm

RC4 has a S-Box S[N], N = 0 to 255 and a secret key, *key*[*l*] where *l* is typically between 5 and 16, used to scramble the S-Box [N]. It has two sequential processes, namely KSA (Key Scheduling Algorithm) and PRGA (Pseudo Random Generation Algorithm) which are stated below in Algorithms 1 and 2 respectively.

| Algorithm 1: KSA | Algorithm 2: PRGA |
|---|---|
| 1.  N = 256;<br>2.  for i = 0 to (N-1)   //Initialization module<br>3.      S[i] = i;         // Identity permutation<br>4.      K[i] = key[i % *l*];<br>5.  end for;<br>6.  j = 0;                           //Storage module<br>7.  for i = 0 to (N-1)<br>8.      j = (j + S[i] + K[i]) % N;<br>9.      swap (S[i], S[j]);<br>10. end for; | 1.  N = 256;<br>2.  i = j = 0;<br>3.  while (TRUE)//Generating Key stream Z<br>4.      i = (i + 1) % N;<br>5.      j = (j + S[i]) % N;<br>6.      swap (S[i] , S[j]);<br>7.      t = (S[i] + S[j]) % N<br>8.      output Z = S[t];<br>9.  end while; |

## 3.  Hardware Implementation of 1-byte 1-clock design

Fig. 1 shows key operations performed by the main processor in conjunction with a coprocessor till the ciphering of the last text character. The hardware for realizing RC4 algorithm comprises of KSA and PRGA units, which are designed in the coprocessor as two independent units, and the XOR operation is designed to be done in the main processor. The central idea of the present embedded system implementation of one RC4 byte in 1-clock is the hardware design of a storage block shown in Fig. 2, which is used in the KSA as well as in the PRGA units. The storage block contains a common S-Box connected to dual select MUX-DEMUX combination and executes the swap operation following line 9 of Algorithm 1 and line 6 of Algorithm 2, in order to update the S-Box. The swap operation in hardware is explained in the following sub-section.

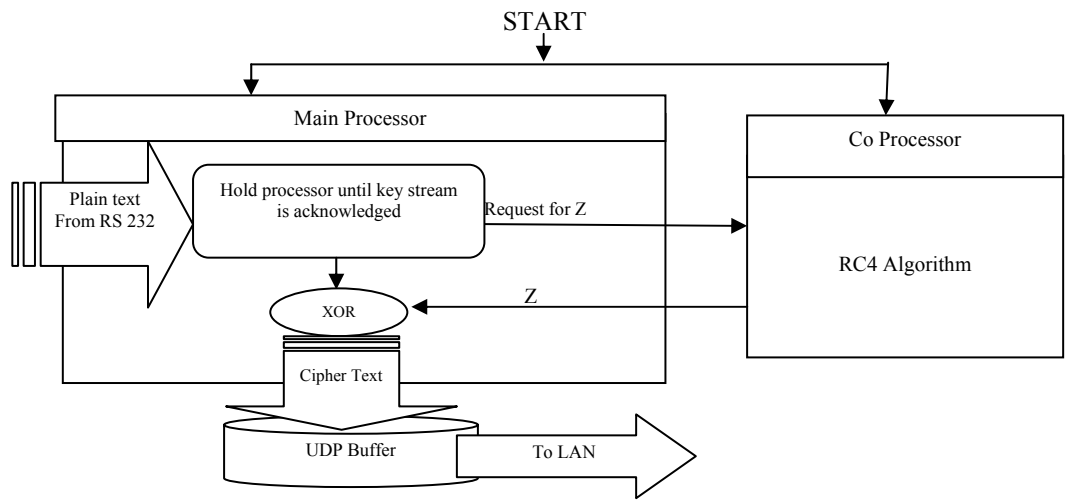

Fig.1: Functioning of the main processor with the coprocessor.

### 3.1 Storage Block Updating the S-Box

The storage block consists of a register bank containing 256 numbers of 8-bit data representing the S-Box, 256:1 MUX, 1:256 DEMUX and 256 D flip-flops. Each of the MUX and DEMUX is so designed, that with 2-select inputs ($i, j$) the two register data (S[$i$], S[$j$]) are operated during the same time. The hardware design of the storage block updating the S-Box is shown in Fig. 2. For swapping, the same S-Box is accessed by the KSA unit with its MUX0-DEMUX0 combination and also by the PRGA unit with its MUX2-DEMUX2 combination and it is also accessed by another MUX3 in PRGA for the generation of the key stream Z. The storage block swaps S[$i$] and S[$j$] and thereby updates the S-Box. For swapping, S[$i$] and S[$j$] ports of MUX are connected to S[$j$] and S[$i$] ports of DEMUX respectively. This storage block has thus 3 input ports ($i, j$ and *CLK*), and 2 inout ports (S[$i$] of MUX, S[$j$] of DEMUX and S[$j$] of MUX, S[$i$] of DEMUX). The storage block of PRGA unit provides 2 output ports from its 2 inout ports which are fed to an adder circuit with MUX3. During the falling edge of a clock pulse, S[$i$] and S[$j$] values corresponding to $i^{th}$ and $j^{th}$ locations of the register bank are read and put on hold to the respective D flip-flops. During the rising edge of the next clock pulse, the S[$i$] and S[$j$] values are transferred to the MUX outputs and instantly passed to the S[$j$] and S[$i$] ports of the DEMUX respectively and in turn are written to the $j^{th}$ and $i^{th}$ locations of the register bank. The updated S-Box is ready during the next falling edge of the same clock pulse, if called for.

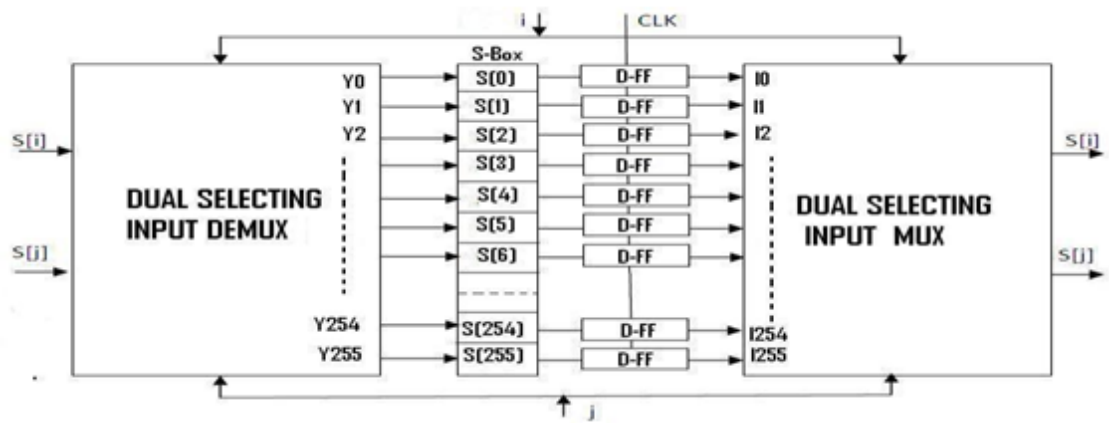

Fig. 2: Storage Block updating the S-Box

### 3.2. Design of the KSA Unit following Algorithm1

Fig. 3 shows a schematic diagram of design of the KSA unit. Initially the S-Box is filled with identity permutation of $i$ whose values change from 0 to 255 as stated in line 3 of initialization module. The $l$-bytes of secret key are stored in the K[256] array as given in line 4. The KSA unit does access its storage block with $i$ being provided by a one round of MOD 256 up counter, providing fixed 256 clock pulses and $j$ being provided by a 3-input adder ($j$, S[$i$], and K[$i$]) following the line 8 of the storage module, where $j$ is clock driven, S[$i$] is MUX0 driven chosen from the S-Box and K[$i$] is MUX1 driven chosen from the K-array. The S-Box is scrambled by the swapping operation stated in line 9 using MUX0-DEMUX0 combinations in the storage block. The KSA operation takes one initial clock and subsequent 256 clock cycles.

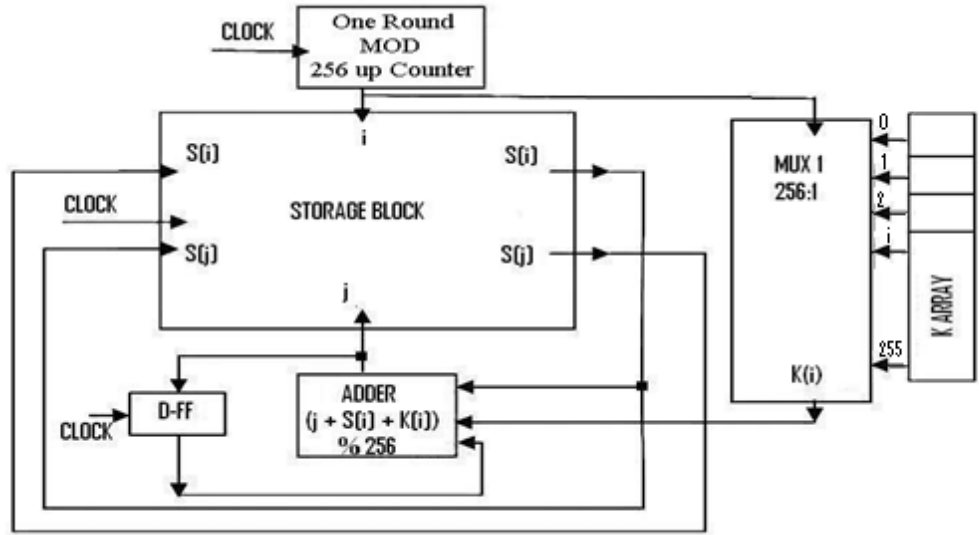

Fig. 3: Schematic Design of the KSA Unit

### 3.3. Design of the PRGA Unit following Algorithm 2

Fig. 4 shows a schematic diagram of the design of the PRGA unit. The PRGA unit does access the storage block with $i$ being provided by a MOD 256 up counter (line 4) and $j$ being given by a 2-input ($j$ and S[$i$]) adder following the line 5 where $j$ is clock driven and S[$i$] is MUX2 driven chosen from the S-Box. With updated *j* and current *i*, the swapping of S[*i*] and S[*j*] is executed following line 6 using MUX2-DEMUX2 combination of the storage block. Following the line 7, S[*i*] and S[*j*] give a value of *t* based on which the key stream *Z* is selected from the S-Box using MUX 3 (line 8).

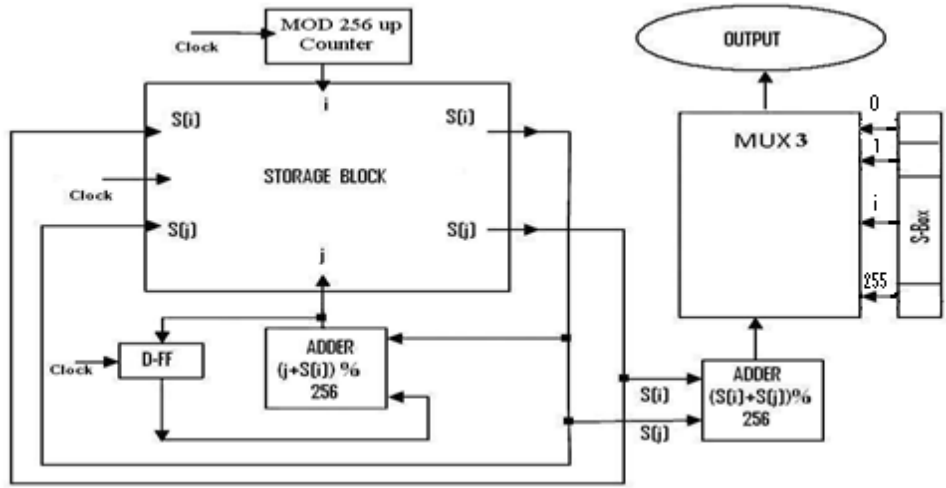

Fig. 4: Schematic Design of the PRGA UNIT

### 3.4    Timing analysis of the PRGA Operation

Let $\phi_i$ denotes the $i^{th}$ clock cycle for $i \geq 0$. It is assumed that the PRGA unit starts when clock cycle is $\phi_0$. It is observed that a signal value gets updated during a falling edge of a clock cycle if it is changed during the rising edge of the previous clock cycle. The symbol ↔ indicates swap operation. The clock-wise PRGA operations are shown in Fig. 5.

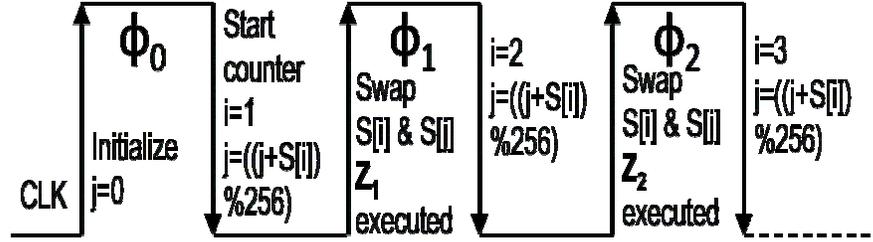

Fig. 5. Clock-wise description of PRGA operation

**Timing Analysis of PRGA**

The MOD 256 up counter shown in Fig. 4 is so designed that *i* starts from 1, goes up to 255 and then it repeats from 0 to 255 for each 256 subsequent clock cycles.

**Rising edge of** $\phi_0$: Initialize $j_0 = 0$.

**Falling edge of** $\phi_0$: Start counter $i_1 = 1$; $j_1 = (j_0 + S[i_1])\%256$; Read S[$i_1$],S[$j_1$].

**Rising edge of** $\phi_1$: S[$i_1$]↔S[$j_1$]; $Z_1 = (S[i_1]+S[j_1])\%256 = 1^{st}$ Key stream.

**Falling edge of** $\phi_1$: $i_2 \rightarrow 2$; $j_2 = (j_1+S[i_2])\%256$; Read S[$i_2$], S[$j_2$].

**Rising edge of** $\phi_2$: S[$i_2$]↔S[$j_2$]; $Z_2 = (S[i_2]+S[j_2])\%256 = 2^{nd}$ Key stream.

**Falling edge of** $\phi_2$: $i_3 \rightarrow 3$; $j_3 = (j_2+S[i_3])\%256$; Read S[$i_3$], S[$j_3$].

The series continues generating successive keys (*Z*'s). If the text characters are *n* and *n* > 254, *i* = 0 after the first round and the clock repeats *(n+1)* times. After generating $Z_n$ during the rising edge of $\phi_n$, PGRA stops. For generating $Z_n$, the PRGA requires *(n+1)* clocks and its throughput per byte is *(1+1/n)*.

### 4.    Experimental Setup and Results

Two Xilinx FPGA Spartan3E (XC3S500e-FG320) boards, each with RC4 encryption and decryption engines separately and shown in Fig 6, are connected through Ethernet ports and to respective hyper terminals (DTE) through RS 232 ports. Two experiments are undertaken. In expt. 1, the whole design was implemented in the Microblaze softcore processor [12] and its code was written using system C. In expt. 2, the RC4 algorithm comprising KSA and PRGA units is designed in a coprocessor using VHDL and the XOR operation is implemented in the main processor using System C. The hardware resource usage for experiments 1 and 2 are narrated in Table 1.

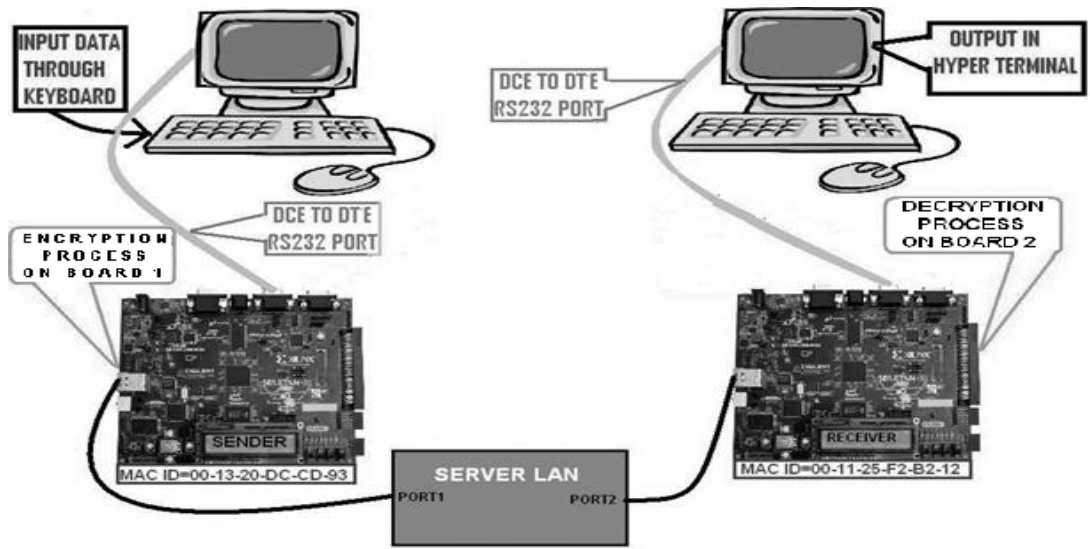

Fig 6 Experimental setup of FPGA based secured data communication

Table 1: Comparison of Resource Usage of two experiments

| Logic Utilization | Hardware Resource Usage | |
|---|---|---|
| | Expt. 1 | Expt. 2 |
| Number of Slice Flip Flops | 2,735 | 6404 |
| Number of 4 input LUTs | 3,039 | 26661 |
| Number of Slices | 5274 | 14448 |

### 4.1 Experimental Results

Table 2 shows the relative comparisons of designs in [1, 2, 3 and 11] with the present design; we show the best values in bold-faced. It can be seen our design is better for most of the design features than that presented in [1] considering large value of n, in spite of the fact that our hardware design is straightforwardly RC4 like and simpler. It is observed that each KSA and PRGA generating a single byte in expt. 1 takes about 430 and 318 clock cycles respectively, while for expt. 2 the corresponding data presented in column 5 of Table 2 below are true.

Table 2: Comparison of various performance metrics with existing designs

| Features | Number of Clock Cycles | | | |
|---|---|---|---|---|
| | Ref.[2,3] | Ref. [11] | Ref. [1] | Our work |
| KSA | 256×3=768 | 3 + 256 = 259 | **1 + 256= 257** | **1 + 256= 257** |
| KSA per byte | 3 | 1 + 3/256 | **1 + 1/256** | **1 + 1/256** |
| PRGA for N-bytes | 3n | 3 + n | 2 + n | 1 + n |
| PRGA per byte | 3 | 1 + 3/n | 1 + 2/n | 1 + 1/n |
| RC4 for N-bytes | 3n+768 | 259+(3+n) | **257+(2+n)** | **257+(1+n)** |
| RC4 per byte | 3+768/n | 1 + 262/n | **1 + 259/n** | **1 + 258/n** |

### 5. Power Optimization

Power optimization study is important in view of its application in emerging embedded technology. In synchronous digital circuits the effective way to reduce the dynamic power dissipation is to dynamically disable the clock in those regions which do not remain active during a specific time of data flow. Since most of the dynamic power consumption in an FPGA is directly related to the toggling of the system clock, temporarily disabling the clock in inactive regions is the most straightforward method of minimizing power consumption. In experiments 1 and 2 stated in Sec. 4 above, there was no clock management process. Fig. 7 shows a schematic diagram of expt. 2 exhibiting KSA and PRGA blocks together.

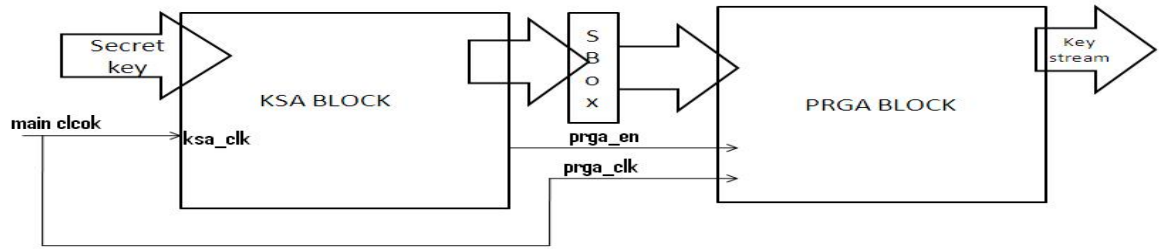

Fig: 7. Circuit Block diagram of Experiment 2.

In RC4 the KSA and PRGA processes are sequential and there is no loss of data if the PRGA block is made active, i.e. prga_en is made '1' from its initial value '0', only when the KSA process finishes all its operations during the first 257 clocks. But both the clocks (ksa_clk and prga_clk) in Fig. 7 are running for the entire computing process. The clock gating circuit incorporated in experiment 2 is shown in Fig. 8.

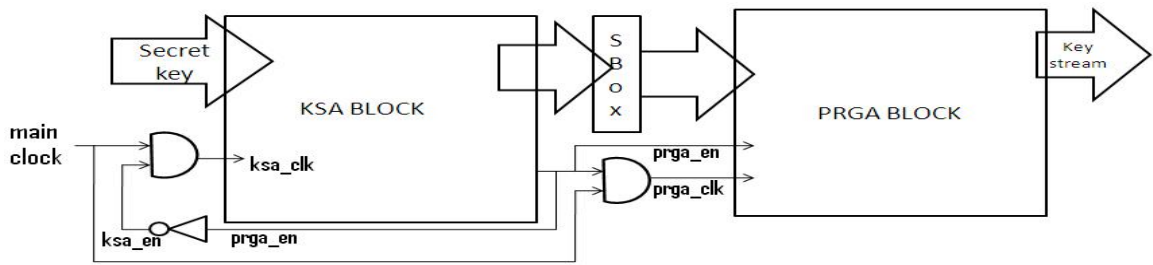

Fig: 8. The incorporation of Clock Gating Circuit in Experiment 2.

The prga_en is first initialized to '0', thereby ksa_en becomes '1' and only the ksa_clk remains active for the first 257 clocks. After the $257^{th}$ clock, prga_en becomes '1', thereby KSA process is instantly disabled and the prga_clk is activativated setting the PRGA block in operation.

Table 3. Results of power on various items

| Power (watt) | Architecture | | |
|---|---|---|---|
| | Behavioral | Structural | Clock Gating |
| Total Power[#] | 1.41665 | 1.18910 | 1.17720 |
| Quiescent Power[#] | 0.96449 | 0.97264 | 0.97080 |
| Dynamic Power[#,*] | 0.45216 | 0.21646 | 0.20640 |
| Clock Power[*] | 0.10280 | 0.11263 | 0.15897 |
| Logic Power[*] | 0.05266 | 0.00545 | 0.00404 |
| IOs Power[*] | 0.00015 | 0.00015 | 0.00002 |
| Signal Power[*,$] | 0.29655 | 0.09823 | 0.04337 |
| Data Signal Power[$] | 0.29628 | 0.09759 | 0.04313 |
| Control Signal Power[$] | 0.00027 | 0.00064 | 0.00024 |

Table 4 shows the power consumed on various items as depicted by the Xilinx X-power [13] analyzer tool doing simulation. It may be noted that the total power is a sum of quiescent and dynamic powers, the dynamic power is a sum of clock, logic, IOs and signal powers and the signal power is a sum of data signal and control signal powers. It is seen from the Table 3 that over the structural design the clock gating technology gives a saving of about 4.6% in dynamic power and about 1% in total power.

## 6. Randomness Tests on RC4 following NIST Statistical Test Suite

Considering the fact that RC4 is very simple, popular and withstood many attacks, it is thought to study the randomness property of its key stream based on 15 statistical tests consolidated by NIST in a Statistical Test Suite [14]. All these statistical tests are undertaken on a sample size of 300 each of which has 1342400 bits produced by RC4. Tests results are shown in Table 3 and in Fig. 9.

Table 4: Number of P-values lying in the given ranges

| Test | 0-.01 | .01-.1 | .1-.2 | .2-.3 | .3-.4 | .4-.5 | .5-.6 | .6-.7 | .7-.8 | .8-.9 | .9-1 |
|---|---|---|---|---|---|---|---|---|---|---|---|
| 1 | 6 | 24 | 29 | 33 | 39 | 26 | 36 | 32 | 24 | 24 | 27 |
| 2 | 1 | 27 | 31 | 31 | 33 | 32 | 31 | 27 | 26 | 25 | 36 |
| 3 | 6 | 24 | 29 | 31 | 32 | 25 | 17 | 35 | 32 | 38 | 31 |
| 4 | 1 | 36 | 39 | 31 | 23 | 29 | 28 | 29 | 21 | 34 | 29 |
| 5 | 3 | 27 | 26 | 34 | 40 | 32 | 26 | 25 | 27 | 31 | 29 |
| 6 | 4 | 19 | 43 | 29 | 26 | 28 | 27 | 34 | 37 | 25 | 28 |
| 7 | 0 | 21 | 35 | 25 | 30 | 26 | 30 | 32 | 26 | 31 | 44 |
| 8 | 4 | 29 | 30 | 28 | 28 | 24 | 37 | 38 | 29 | 25 | 28 |
| 9 | 2 | 23 | 31 | 24 | 25 | 42 | 31 | 29 | 28 | 33 | 32 |
| 10 | 4 | 29 | 26 | 34 | 29 | 39 | 22 | 28 | 23 | 32 | 34 |
| 11 | 5 | 61 | 71 | 62 | 56 | 49 | 51 | 65 | 62 | 62 | 56 |
| 12 | 3 | 26 | 39 | 31 | 30 | 27 | 26 | 31 | 31 | 29 | 27 |
| 13 | 7 | 55 | 62 | 61 | 44 | 57 | 71 | 55 | 70 | 58 | 60 |
| 14 | 31 | 217 | 231 | 250 | 245 | 264 | 249 | 236 | 252 | 204 | 215 |
| 15 | 72 | 477 | 543 | 572 | 557 | 561 | 531 | 567 | 503 | 522 | 495 |

The P-value in the NIST tests is the probability value indicating the degree of non-randomness – the lesser is the P-value, the higher is its degree of non-randomness. For a particular bit sequence, if its value for a particular test is less than 0.01, the sequence is considered to be completely non-random. Considering all the P-values for a particular test to be undertaken on samples of size (N) greater than 100, one can define a parameter $P_{pop}$ as proportion of passing of P-values. The theoretical statistical estimate of acceptable $P_{pop}$ is $0.99 \pm R$, where R is inversely proportion to the square root of N, the larger the sample size, the smaller the value of R. Considering 300 samples of RC4 key bits sequences obtained from 300 different keys, the value of R is calculated approximately as 0.01. The observed $P_{pop}$ is the relative number of P-values lying above 0.01 to all the P-values. From the said statistical consideration, all the 300 samples of RC4 key bit sequences is observed to pass all the 15 tests although 149 P-values are found to fail among 12300 (=41x300) P-values. It is to be noted that for a particular sample, test nos. 1 – 10 and 12 have one P-value each, while test nos. 11 and 13 have 2 P-values each, 14 has 8 P-values and 15 has 18 P-values – altogether 41 P-values. Table 4 shows number of P-values lying in 11 ranges between 0 and 1 for all 15 tests. Fig. 8 depicts the observed proportion of passing (Y-axis) for all tests (X-axis). Among the 15 tests, the lowest observed $P_{pop}$ is 0.98 for tests 1 and 3, while the highest one is 1.00 for test 7. The observed $P_{pop}$ for all the 15 tests are shown in Fig. 9.

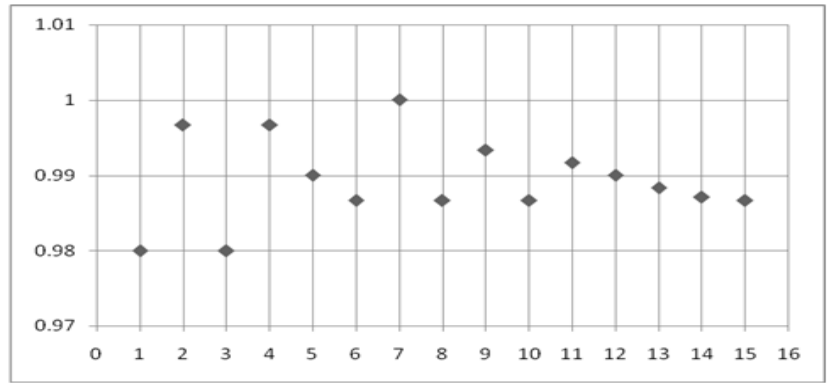

Fig. 9. Observed Proportion of passing of RC4

The P-value of P-values (POP) for a particular test is another parameter whose value is calculated based on Table 4 following a statistical methodology mentioned by NIST [14]. The distribution of P-values for a particular test undertaken on all the samples can be considered uniform, if its POP is greater than 0.0001. From Table 5 it is seen that the POP of all the 15 tests are above 1e-4 and one can conclude that P-values of all the 15 tests are uniformly distributed. It is also seen that the POP value is most for test 2 and least for test 14 exhibiting the fact that test 2 produces most uniformly distributed 300 P-values and test 14, the least – although both exhibit uniform distribution of P-values. From the histograms of Figs. 10 (a) and (b), one can very easily visualize the uniformity distribution of P-values for tests 2 and 14. In both the histograms there are 10 columns: first column indicates the number of P-values lying between 0 and 0.1; second column indicates that between 0.1 and 0.2, so on and so forth. The column height indicates the frequency counts of P-values for each histogram as shown in Table 4. One can thus conclude that according to NIST Statistical Test Suite the RC4 key bit sequences can be considered to be fairly random.

Table 5: Status for Proportion of Passing and Uniformity of Distribution

| Test | Expected Proportion | Observed Proportion | Status for Proportion of Passing | P-value of P-values (POP) | Status for Uniform / Non-uniform Distribution |
|---|---|---|---|---|---|
| 1 | 0.972766 | 0.980000 | Successful | 5.744434e-01 | Uniform |
| 2 | 0.972766 | 0.996667 | Successful | 9.393588e-01 | Uniform |
| 3 | 0.972766 | 0.980000 | Successful | 3.665526e-01 | Uniform |
| 4 | 0.972766 | 0.996667 | Successful | 3.949802e-01 | Uniform |
| 5 | 0.972766 | 0.990000 | Successful | 7.127007e-01 | Uniform |
| 6 | 0.972766 | 0.986667 | Successful | 2.490301e-01 | Uniform |
| 7 | 0.972766 | 1.000000 | Successful | 2.056983e-01 | Uniform |
| 8 | 0.972766 | 0.986667 | Successful | 6.852543e-01 | Uniform |
| 9 | 0.972766 | 0.993333 | Successful | 5.004468e-01 | Uniform |
| 10 | 0.972766 | 0.986667 | Successful | 4.681345e-01 | Uniform |
| 11 | 0.977814 | 0.991667 | Successful | 6.228382e-01 | Uniform |
| 12 | 0.972766 | 0.990000 | Successful | 9.113979e-01 | Uniform |
| 13 | 0.977814 | 0.988333 | Successful | 4.617824e-01 | Uniform |
| 14 | 0.983907 | 0.987083 | Successful | 2.002723e-01 | Uniform |
| 15 | 0.985938 | 0.986667 | Successful | 2.245991e-01 | Uniform |

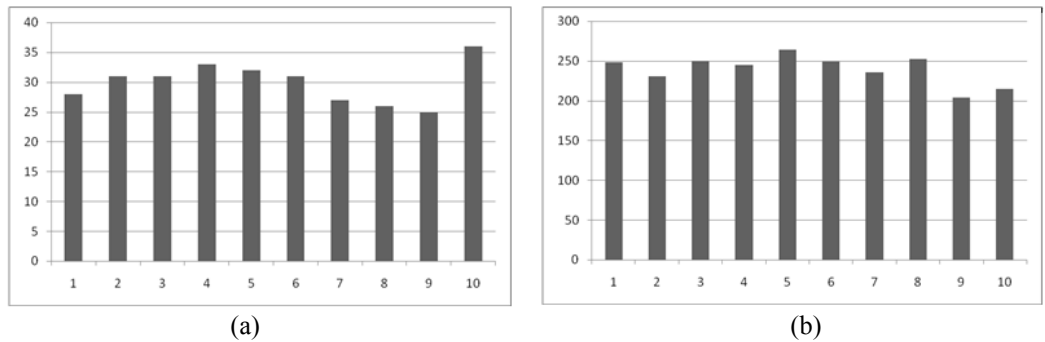

Fig.. 10. Uniformity Distribution of P-values of (a) Test 2 and (b) Test 14.

## 7. Conclusion

The proposed 1-byte-1-clock RC4 design in FPGA is a coprocessor based design functioning in parallel with a main processor. The encryption engine of the design implemented in one board successfully communicates through its Ethernet port to another board containing the decryption engine. The present 1-byte 1-clock processing exploits conventional VHDL features and circuit-wise it is much simpler than the processing of 2-bytes together in 2 clocks [1], leading to a throughput little better than that presented in [1, 11]. The clock gating technology incorporated in the structural design is found to reduce dynamic power by about 5%. From the statistical randomness studies, RC4 is found to be producing reasonably fair random key bit sequences.